**Title: Aperiodic MEG abnormality in patients with focal to bilateral tonic-clonic seizures**

**Authors :** Kirandeep Kaur [1,2,3,4], Jonathan J Horsley[1], Csaba Kozma[1], Gerard R Hall[1], Thomas W Owen[1], Yujiang Wang[1,5,6,7], Guarav Singh[8], Sarat P Chandra[8,9], Manjari Tripathi[4,8], Peter N Taylor[1,5,6,7,*]

1. Computational Neuroscience, Neurology, and Psychiatry Lab (www.cnnp-lab.com), ICOS Group, School of Computing, Newcastle University, Newcastle upon Tyne, United Kingdom
2. Institute of Health and Neurodevelopment, Aston University, Birmingham, United Kingdom
3. Wellcome Centre for Human Neuroimaging, University College London, United Kingdom
4. Department of Neurology, All India Institute of Medical Sciences, New Delhi India
5. UCL Queen Square Institute of Neurology, London, United Kingdom
6. National Hospital for Neurology and Neurosurgery, London, United Kingdom
7. Faculty of Medical Sciences, Newcastle University, Newcastle upon Tyne, United Kingdom
8. MEG Facility, National Brain Research Centre, Manesar, India
9. Department of Neurosurgery, All India Institute of Medical Sciences, New Delhi India

* peter.taylor@newcastle.ac.uk

**Abstract:**

**Background:** Aperiodic activity is a physiologically distinct component of the electrophysiological power spectrum. It is suggested to reflect the balance of excitation and inhibition in the brain, within selected frequency bands. However, the impact of recurrent seizures on aperiodic activity remains unknown, particularly in patients with severe bilateral seizures. Here, we hypothesised greater aperiodic abnormality in the epileptogenic zone, in patients with focal to bilateral tonic clonic (FBTC) seizures, and earlier age of seizure onset.

**Methods:** Pre-operative magnetoencephalography (MEG) recordings were acquired from 36 patients who achieved complete seizure freedom (Engel I outcome) post-surgical resection. A normative whole brain map of the aperiodic exponent was computed by averaging across subjects for each region in the hemisphere contralateral to the side of resection. Selected regions of interest were then tested for abnormality using deviations from the normative map in terms of z-scores. Resection masks drawn from postoperative structural imaging were used as an approximation of the epileptogenic zone.

**Results:** Patients with FBTC seizures had greater abnormality compared to patients with focal onset seizures alone in the resection volume (p=0.003, area under the ROC curve = 0.78 ). Earlier age of seizure onset was correlated with greater abnormality of the aperiodic exponent in the resection volume (correlation coefficient = -0.3, p= 0.04)) as well as the whole cortex (rho = -0.33, p=0.03). The abnormality of the aperiodic exponent did not significantly differ between the resected and non-resected regions of the brain.

**Conclusion:** Abnormalities in aperiodic components relate to important clinical characteristics such as severity and age of seizure onset. This suggests the potential use of the aperiodic band power component as a marker for severity of epilepsy.

# Introduction

The spectral activity of magnetoencephalography (MEG) consists of rhythmic oscillations which appear as peaks on a power spectrum and an aperiodic component which appears as a slope at the base of this oscillatory activity (Donoghue et al., 2020). The aperiodic component is hypothesised to represent activity from neuronal populations which are not oscillating (Freeman & Zhai, 2009; He, 2014; Miller et al., 2009). This activity follows the 1/f power law, wherein the power declines exponentially as a function of the frequency (He, 2014; Miller et al., 2009). The rate at which this power declines is measured by the aperiodic exponent which captures the steepness, or 'flatness' of the 1/f slope. The shift in power across all frequencies is measured by the aperiodic offset (Donoghue et al., 2020). Together, the aperiodic exponent and offset constitute the two parameters used to model the aperiodic activity of the power spectrum.

Conventionally, the aperiodic component was often considered to be the 'noise' in the power spectrum, and not of any physiological importance (Donoghue et al., 2020; He et al., 2010). However, recent findings have generated immense interest in the aperiodic activity after it was found to alter with natural ageing (Hill et al., 2022; Merkin et al., 2023), psychiatric disorders (Ostlund et al., 2021; Peterson et al., 2023) and other clinical conditions including Parkinson's and Tourette Syndrome (Adelhöfer et al., 2021; Wiest et al., 2023; Wilkinson & Nelson, 2021).

Previous research in epilepsy has largely focused on the rhythmic component of the power spectrum and its relevance to the mechanistic understanding of seizures and surgical outcomes. In contrast, the underlying aperiodic activity is relatively unexplored. However, some studies demonstrated potential utility of aperiodic activity in the field of epilepsy research (Coa et al., 2022; Kundu et al., 2023; Pani et al., 2021; van Heumen et al., 2021). Specifically, in two recent case studies, the aperiodic exponent varied temporally and spatially, serving as an independent marker of epileptogenicity and seizure onset zone respectively (Kundu et al., 2023; van Heumen et al., 2021). Those case studies did not include patient group analyses, and the utility of aperiodic exponent in epilepsy research based on larger scale data remains unknown.

In addition, empirical evidence suggests that increased aperiodic exponent is correlated with increased synchrony of neuronal populations (Freeman & Zhai, 2009). Computational models also indicate that the aperiodic exponent may estimate the balance of excitation and inhibition in underlying neuronal circuits between 30-70Hz (Gao et al., 2017; He et al., 2010). Interestingly, seizure generation is also characterised by excessive synchronisation of neuronal populations due to an imbalance towards excitatory activity in the brain (Bromfield et al., 2006; van van Hugte et al., 2023).

Taken together, research suggests the aperiodic activity may vary with epileptogenic activity. However, to date, the following questions remain unanswered - Do more severe forms of epilepsy, such as presence of focal to bilateral tonic clonic (FBTC) seizures (Sinha et al., 2021) show greater abnormality of the aperiodic component? Does the abnormality of aperiodic component correlate with other clinical variables such as age of seizure onset? Is the aperiodic component differentially expressed between the epileptogenic zone and other regions of the brain?

To explore these questions, we extracted the aperiodic exponent from preoperative MEG records of 36 patients who underwent subsequent surgical resection and had an Engel I outcome (complete seizure remission). We hypothesised that (i) patients with more secondary generalised seizures (FBTC seizures) will have greater abnormality of the aperiodic exponent as compared to those who present with focal seizures alone. In addition, we hypothesised that greater abnormality in the aperiodic exponent positively correlated with (iii) earlier age of onset in epilepsy (iii) resected tissues of the cortex.

# Methods

*Patient characteristics*

Preoperative MEG recordings and MRI were acquired for all patients. Postoperative MRI or CT scans were also acquired within 24 hours of surgery. A clinical history of 36 patients was retrieved from the patient repository of All India Institute of Medical Sciences, New Delhi, India. All patients had complete seizure remission (Engel I outcome) for at least three years post-surgery. Clinical and demographic characteristics are described in Table I.

**Table I: Clinical and demographic characteristics of patients**

| Clinical and demographic characteristics | |
| --- | --- |
| Age [Mean±SD] | 15.47 ± 8.45 |
| Duration of epilepsy in years [Mean±SD] | 8.29 ± 6.08 |
| Age at seizure onset in years [Mean±SD] | 7.18 ± 7.55 |
| Seizure frequency/ week [Median (IQR)] | 7 (1.44 - 19.25) |
| Sex<br>• Female (%)<br>• Male (%) | 14 (38.9)<br>22 (61.1) |
| Multilobar resection<br>• No (%)<br>• Yes (%) | 25 (69.4)<br>11 (30.6) |
| Presence of FBTC seizures<br>• No (%)<br>• Yes (%) | 11 (33.3)<br>25 |

**MRI processing and drawing of resection masks**

The resection masks were generated in two steps. Firstly, the preoperative volumetric T1 MRIs were preprocessed to generate cortical parcellations using FreeSurfer (Fischl, 2012). The

Lausanne parcellation scheme used for this study comprises 114 neocortical regions of interest (ROIs) (Hagmann et al., 2008).

In the second step, postoperative CT/MRI were linearly registered to the preprocessed preoperative MRIs using the FSL FLIRT tool (Jenkinson et al., 2002; Jenkinson & Smith, 2001). As an additional step, postoperative CT scans were contrast enhanced for better visibility of the resection cavity using the software ITK-SNAP (Yushkevich et al., 2006). The co-registered postoperative scans were then overlaid on preoperative scans using the software FSLView. Masks were manually drawn to fill the resection cavity on every alternative sequence of the 'orig.mgz' file of the FreeSurfer processed MRI, and subsequently smoothed using the 'fslmaths' toolkit (Figure 1).

An ROI was considered to be resected if there was a difference of at least 10% between the pre and post-operative volume. ROIs with <10% change in volume were retained as non-resected regions.

**MEG acquisition and pre-processing**
MEG data were acquired for all patients preoperatively using the 306 channel Elekta Neuromag system in an eyes-closed, resting state. Details of MEG acquisition protocols for this set of patients have been described previously (Kaur et al., 2021; Tripathi et al., 2021). Raw MEG recordings (downsampled to 500Hz) were pre-processed, cleaned and analysed using Brainstorm software (Tadel et al., 2011).

MEG recordings were co-registered with the patients' Freesurfer processed T1 MRI using anatomical fiducials (nasion, right preauricular, left preauricular, anterior commissure, posterior commissure, interhemispheric point). Data was then notch filtered at 50Hz using a second order IIR (infinite impulse response) notch filter, and then band-pass filtered between 1-100Hz. Cardiac and ocular artefacts were removed using the signal space projection (SSP) spatial decomposition method. Following filtering and artefact removal, the pre-processed MEG file was source modelled using the sLORETA approach and a head model constituted of overlapping spheres. This source model was downsampled by averaging into 114 neocortical regions using the Lausanne parcellation scheme described previously. This resulted in the generation of one time series per region of interest, which was then exported for analysis of its aperiodic activity (Figure 1).

Power spectrum density (PSD) estimates from the time series were generated using the Brainstorm software using the band-pass settings described above. Welch's method was used to compute the PSD, the window length was restricted to two seconds and the window overlap period was 50%. Artefacts, likely resulting from interference between head position indicator (HPI) coils at 7Hz and subsequent harmonics, were removed using a custom MATLAB script. The script identifies peaks at these harmonic frequencies and interpolates the PSD to reduce their impact.

**Computing aperiodic components**

The aperiodic component was computed using the specparam (FOOOF) toolbox (Donoghue et al., 2020) and its MATLAB wrapper. Aperiodic parameters were computed for each ROI time series between the frequency range of 1 to 47.5 Hz, with the maximum number of peaks set to two.

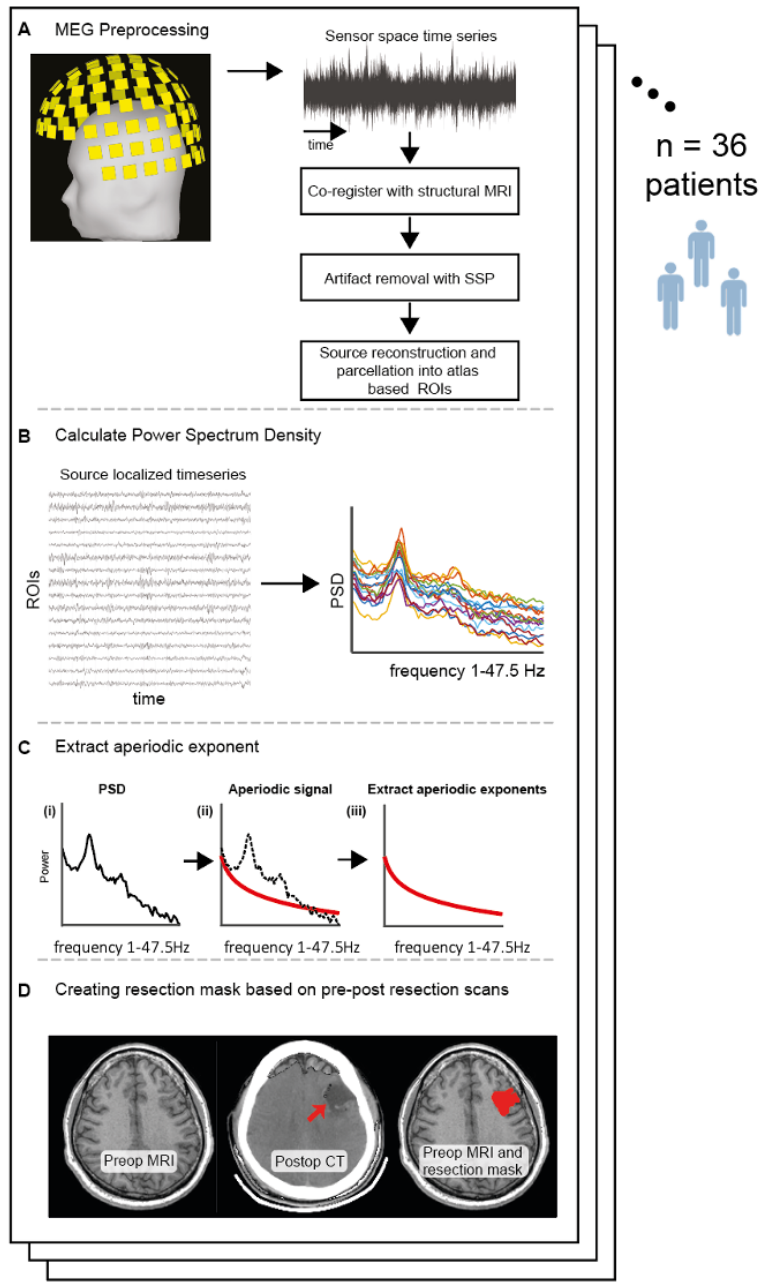

**Figure 1. (A)** Preoperative resting state MEG scans and structural MRI were retrieved for 36 patients who underwent surgical resection for drug resistant epilepsy (DRE), and were seizure free for at least 3 years. The preoperative structural MRI were parcellated using FreeSurfer and co-registered with the MEG recordings. This MEG sensor time-series data was filtered, cleaned using SSP and source localised to generate scout (ROI) time-series in Brainstorm software **(B)** Power spectral density (PSD) of the source localised time series was computed **(C)** Aperiodic exponent was extracted from the PSD using the specparam (FOOOF) toolbox **(D)** A resection mask for each patient was generated by co-registering the postoperative MRI/CT to the preoperative MRI and manually filling the resection cavity.

**Abnormality mapping of aperiodic exponent**

The FOOOF algorithm generates the aperiodic exponent (1/f slope) of the PSD through an iterative process (Donoghue et al., 2020). In this study, we investigated this aperiodic exponent (Figure 1).

We mapped the abnormality in the aperiodic exponent using the following approach- in the first step, the hemisphere contralateral to the region of resection was presumed to contain 'normative' values of the aperiodic activity and these were then averaged across patients for each ROI to create a normative distribution of the aperiodic exponent for the whole cortex. In the second step, z-score values were derived for every ROI in each patient, using the 'normative map' as the reference distribution. A z-score value indicates how many standard deviations away a given value is from the mean. In other words, the greater the absolute z-score of an ROI in a patient, the more it differs from the mean normative value of that region. In the third step, these z-scores were averaged across four regions for all subjects: a) the resection volume b) the ipsilateral hemisphere c) the contralateral hemisphere d) the whole cortex.

**Statistical Analysis**

We used a Wilcoxon signed-rank test to assess if the mean abnormality of the resection volume differed between patients who present with FBTC seizures as compared to those who do not. Correlation analysis was conducted to evaluate if age of seizure onset was associated with greater mean abnormality of aperiodic activity.

Our final objective was to assess if the mean resection volume had greater abnormality of aperiodic exponent as compared to the non-resected regions of the brain. The $D_{RS}$ measure, which stands for the distinguishability between the resected and spared brain regions, was used to compare the rank order of the abnormality between the two (Owen et al., 2023). The $D_{RS}$ varies between 0 to 1, and values close to 0 indicate that the resected regions are more abnormal than the non-resected, while values close to 1 suggest the opposite. $D_{RS}$ of 0.5 indicates chance, i.e. there is no difference between the resected and non-resected regions in terms of abnormality.

## Results

In this study we assessed if patients with focal to bilateral tonic-clonic seizures (FBTCS+) have more abnormal aperiodic exponents compared to patients who do not have focal to bilateral tonic-clonic seizures (FBTCS-). Secondly, we tested the association of aperiodic abnormalities with the age of seizure onset. Lastly, we compared the abnormality of the aperioidic exponent between the resected and spared tissues of the brain.

**Aperiodic exponent abnormality in patients with FBTC seizures**

The resection volume was more abnormal pre-operatively in FBTCS+ patients than those with focal onset seizures alone (p=0.003) (Figure 2A).

The effect size of the difference in abnormality between patients who had FBTC seizures vs focal onset seizures alone was large (Cohen's d = 0.9, 95% confidence interval [0.13, 1.67]). The area under this ROC curve (AUC) represents the accuracy with which aperiodic exponent can differentiate between patients who have FBTC seizures and those who don't. In our study AUC was found to be 0.78, on a scale from 0-1 (Figure 2B).

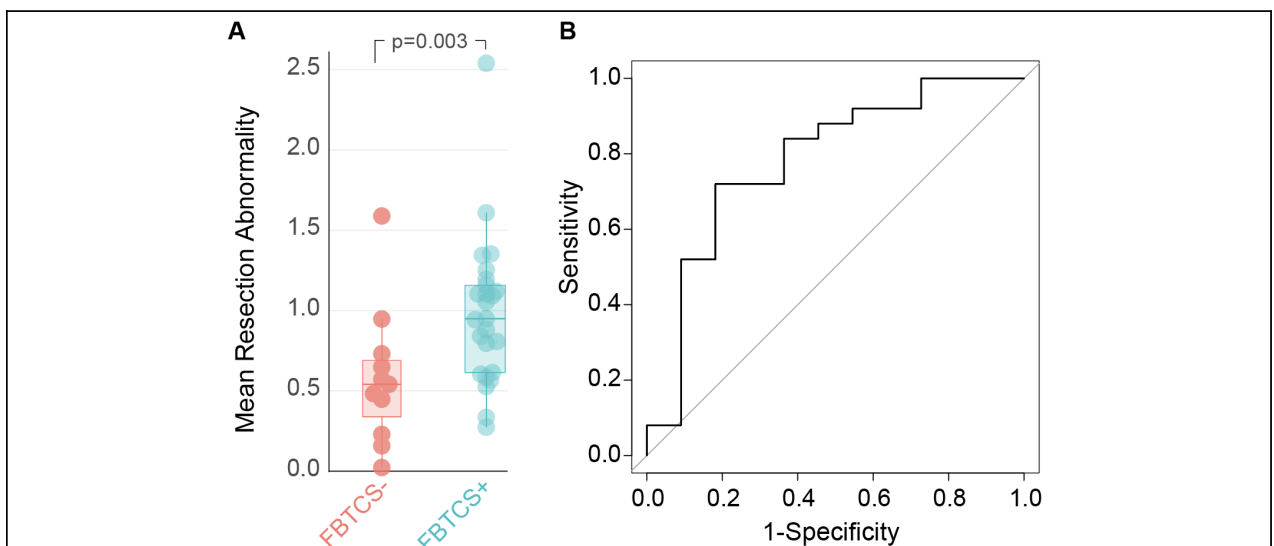

**Figure 2: Patients with focal to bilateral tonic-clonic (FBTC) seizures have greater abnormality in the resected volume as compared to patients who present with focal onset seizures alone. (A)** The distribution of the mean abnormality of the resection volume in patients who presented with FBTC seizures and those who did not. The upper and lower bars of the box plot represent the first and third quartile of the distribution, and the thick central line represents the median. Every data point represents one individual

patient. **(B)** Area under the receiver operating characteristic curve for aperiodic exponent as a classifier of the presence of FBTC seizures.

**Association of aperiodic exponent abnormality with age of seizure onset**

We investigated if the magnitude of aperiodic exponent abnormality was associated with age of seizure onset. In our cohort, earlier age of seizure onset correlated with mean aperiodic abnormality of the whole cortex (rho = -0.33, p=0.03), ipsilateral hemisphere (rho = -0.32, p=0.03) and the resection volume (rho = -0.3, p=0.04) (Figure 3).

These results suggest that the aperiodic spectrum accumulates greater abnormalities in patients who have seizures earlier in life. These abnormalities are not restricted to the resected regions alone and affect the whole cortex.

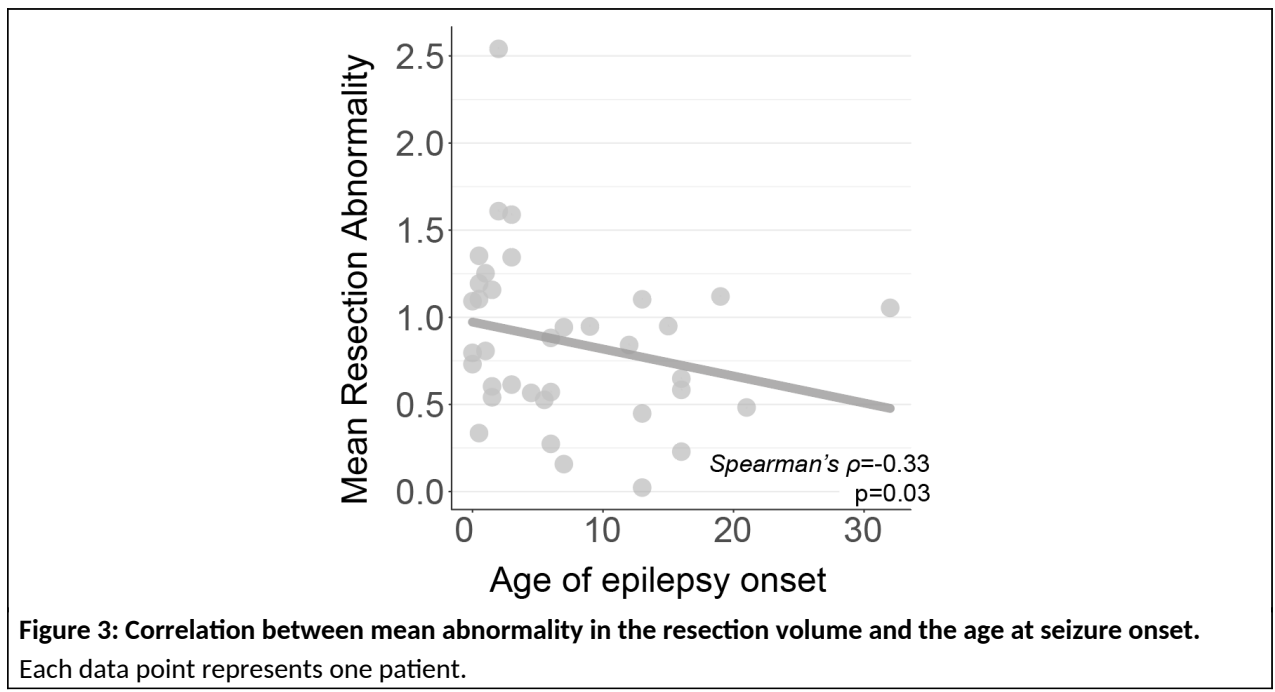

**Figure 3: Correlation between mean abnormality in the resection volume and the age at seizure onset.** Each data point represents one patient.

**Abnormality of aperiodic exponent in resected and non-resected regions of the brain**

We computed $D_{RS}$ scores for all 36 patients and used a single sample Wilcoxon signed-rank test to assess if the median value of these scores differed significantly from 0.5. The median score of $D_{RS}$ was 0.47, and this value did not significantly differ from 0.5 (p= 0.43). This suggests that resected areas were *not* more abnormal than spared areas in our cohort, in their aperiodic exponent.

In summary, these results suggest that the aperiodic spectrum shows greater abnormality in patients who present with FBTC seizures as well as those who have an earlier age of epilepsy onset.

## Discussion

In this study we extracted the aperiodic exponent from resting preoperative MEG scans of patients who had complete seizure remission following epilepsy surgery. We found that the aperiodic (1/f) exponent of the power spectrum in the presumed epileptogenic area was more abnormal in patients who have FBTC seizures and those who have onset of seizures from an early age. Furthermore, in patients with earlier age of onset, this association with abnormal aperiodic activity extended to the whole cortex, including the hemispheres ipsilateral and contralateral to the site of resection.

In the first analysis, we found that patients who present with FBTC seizures have greater abnormality of the aperiodic exponent in the assumed epileptogenic zone as compared to those patients who do not. Since the aperiodic exponent is hypothesised to be a proxy marker of E/I in the cortex (Ahmad et al., 2022), our results indicate a greater disruption of E/I balance in FBTC seizures as compared to focal seizures alone. FBTC seizures are known to originate in one part of the cortex, and then spread bilaterally to involve both hemispheres (Fisher, Cross, D'Souza, et al., 2017; Fisher, Cross, French, et al., 2017). These are considered to be the one of the most severe seizure types as they carry the highest risk of seizure related fatal injury, pharmacoresistance and sudden death in epilepsy (SUDEP) (Devinsky et al., 2016; Lawn et al., 2004, Baud et al., 2015, Bone et al, 2012, Chang et al, 2023). There exist competing theories regarding the pathways underlying the spread of FBTC seizures to the contralateral hemisphere. Initially, it was proposed that FBTC seizures recruit canonical thalamocortical circuits for propagation, but latest findings suggest that in addition, these seizures also propagate via the striatum, globus pallidus and corpus callosum (Brodovskaya & Kapur, 2019; He et al., 2020; Unterberger et al., 2016). Here, the basal ganglia were proposed to act as a braking mechanism, which inhibited the spread of seizures to the contralateral side (He et al., 2020; Vuong & Devergnas, 2018). In line with this hypothesis, He et al found increased thalamus mediated cortico-cortical interactions, and reduced basal ganglia-thalamus interactions in resting state functional MRI in patients with FBTC seizures (He et al., 2020).

Sinha et al found that patients with FBTC seizures present with greater widespread abnormalities of whole brain structural networks as compared to patients who have focal seizures alone (Sinha et al., 2021). In addition, Maher et al reported that white matter disruptions in subcortical tracts- including striatal and thalamic occipital tracts, in patients with FBTC seizures may render these pathways more conducive to seizure propagation in the adjoining hemisphere (Maher et al., 2022). Our study lends further support to the idea that FBTC seizures are characterised by greater, and more widespread, abnormality of underlying brain circuits as compared to focal seizures.

In our second analysis, we observed a significant correlation between earlier age of epilepsy onset and aperiodic abnormality. Current literature posits that earlier age of epilepsy onset is prognostic of poor outcomes in terms of seizure freedom and normal development (Berg et al., 2019; Nickels, 2019). Park et al reported that age of epilepsy onset was negatively correlated with duration of symptoms, i.e. older age at onset was associated with a greater probability of seizure freedom (Park & Lee, 2020). Aguglia et al reported similar findings in patients with sporadic non-lesional epilepsy (Aguglia et al., 2011). From these findings we conclude that earlier age of seizure onset is associated with less favourable prognosis as well as greater aperiodic abnormality. Taken together, these results suggest that more severe epilepsy is accompanied by greater abnormality of the aperiodic exponent.

Lastly, we found that resected regions of the brain were not significantly abnormal in the aperiodic activity as compared to the non-resected brain regions. These differences could not be detected even when the resection volume was compared to the contralateral hemisphere. These results are in agreement with the case report by Heumen et al which did not find significant differences in aperiodic activity between the resected and non-resected brain regions during the interictal period(van Heumen et al., 2021). However, in the same report, the authors noted a difference in aperiodic abnormality between the seizure onset zone and control regions of the brain shortly before the onset of a seizure. While it is entirely possible that the resected and non-resected brain regions may not have any physiological differences in the aperiodic activity, in our study these results may have been an outcome of the manner in which the normative map for comparison was computed. In the absence of healthy controls, we presumed the contralateral hemisphere to represent a *normal distribution* of the aperiodic activity and serve as a control in comparison to the disrupted ipsilateral hemisphere. However, this assumption may have bias as the epileptogenic networks are known to extend bilaterally (Bernasconi, 2017). Therefore, it is possible that the contralateral hemisphere itself contained some abnormal activity. This may have shifted the normative map itself towards greater abnormality, thus reducing the abnormality gap between the resected and non-resected tissues. This result is also in agreement with Kozma et al (2023), who showed that both periodic *and* aperiodic components were required to best localise epileptogenic tissues.

This study had several strengths and limitations. A key strength is that all patients had an Engel I (seizure-free) outcome for at least three years post-surgery. This raises our confidence that the resected brain tissue was indeed epileptogenic. A further strength is that the resection volume was manually delineated using postoperative scans. For many research studies this would not be possible due to lack of available data. A limitation of our data is the absence of healthy controls, which led to our use of contralateral regions for normative map generation. This means our findings are likely conservative, since deviation from truly healthy brain recordings would be expected to be greater, as recently shown (Owen et al., 2023). Nonetheless, the approach of using apparently normal regions in patients was shown to be useful in intracranial EEG (Bernabei et al., 2022; Taylor et al., 2022), and our evidence here further supports this approach.

Collective evidence suggests that aperiodic activity is an independent and physiologically relevant component of the power spectrum (Donoghue et al., 2020; He, 2014; He et al., 2010). Epilepsy is characterised by hyper-synchronisation of neural circuits, and the aperiodic exponent is conjectured to provide a non-invasive estimate of the same between the 30-70Hz frequency bands (Freeman & Zhai, 2009; Voytek & Knight, 2015). In addition, preliminary evidence suggests that a decrease in aperiodic components is associated with sustained seizure freedom post-surgery (Kundu et al., 2023). Lastly, the analysis of aperiodic activity has a major advantage that it is highly conserved in individuals (Demuru & Fraschini, 2020). In summary, a synthesis of previous findings and the results of this study suggest that the aperiodic component of the power spectrum may serve as a stable marker of disease progression. However, further research is needed to validate this.

In conclusion, we observed that aperiodic abnormality was higher in epilepsy with wider networks and bilateral tonic clonic spread. Our study provides further support to the hypothesis that aperiodic activity in epilepsy can serve as a marker of epileptogenicity.

## **Funding**

Magnetoencephalography (MEG) Resource Facility (BT/MED/122/ SP24580/2018) from the Department of Biotechnology, Ministry of Science & Technology, Govt. of India to All India Institute of Medical Sciences (AIIMS), New Delhi and the National Brain Research Center. C.K. is supported by Epilepsy Research UK, P.N.T. and Y.W. are both supported by UKRI Future Leaders Fellowships (MR/T04294X/1, MR/V026569/1). J.J.H. is supported by the Centre for Doctoral Training in Cloud Computing for Big Data (EP/L015358/1).

## **Acknowledgements**